\documentstyle{mn}

\input epsf

\begin{document}

\journal{Preprint-98}

\title{Higher order moments of the density field in a parameterized
sequence of non-gaussian theories}

\author[M. White]{Martin White\\
Department of Astronomy, Harvard University\\
60 Garden St, Cambridge MA 02138, USA\\
(mwhite@cfa.harvard.edu)}
\pubyear{1999}

\maketitle

\begin{abstract}
We calculate the higher order moments in a sequence of models where
the initial density fluctuations are drawn from a $\chi^2_\nu$ distribution
with a power-law power spectrum.  For large values of $\nu$ the distribution
is approximately gaussian, and we reproduce the values known from
perturbation theory.  As $\nu$ is lowered the distribution becomes
progressively more non-gaussian, approximating models with rare,
high-amplitude peaks.  The limit $\nu=1$ is a realization of recently
proposed isocurvature models for producing early galaxy formation where
the density perturbations are quadratic in a gaussian field.
\end{abstract}


\section{Introduction}

The standard paradigm for the formation of large scale structure is that
quantum fluctuations during an inflationary epoch seeded initially small,
gaussian fluctuations in density, which grew through the action of
gravitational instability in a universe whose dominant constituent is cold
dark matter (CDM).
This theory has proved very predictive and agrees with a wide range of
observational data, however several of the assumptions are difficult to
test with high precision.
The assumption that the initial fluctuations have a gaussian distribution
holds more generally than just in the inflationary CDM model, as
the central limit theorem implies that fluctuations emerging from many
uncorrelated, random processes will be nearly gaussian.  Though it is
indeed a plausible assumption, and a prediction in the simplest
inflationary models, it has received only limited observational support
through measurements of the CMB \cite{KBBGHSW,Heavens}
and large-scale structure \cite{BSDFYH,Gaz,NusDekYah,FKP,StiPea,Colley}.
In the latter case the situation is made more difficult by the action
of gravity which turns an initially gaussian random field into a non-gaussian
field once the modes become non-linear.

For gaussian fluctuations the only non-trivial moment is the 2-point
correlation function, $\xi(r)$, or its Fourier Transform the power spectrum
$P(k)$.
If the fluctuations are non-gaussian the higher-order moments of the field
carry additional information.
The evolution of the higher order moments induced by gravitational instability
in an initially gaussian random field has been studied extensively
(see Strauss \& Willick~\shortcite{StrWil} for a review).
Of particular interest are the $S_N$, defined in terms of the volume
averaged correlation functions
(cumulants of the probability distribution function)
\begin{equation}
  \bar{\xi}_N \equiv {1\over V^N} \int_V d^3r_1\,d^3r_2\,\cdots\,d^3r_N
  \ \xi({\bf r}_1,\cdots,{\bf r}_N)
\end{equation}
as
\begin{equation}
\bar{\xi}_N=S_N\bar{\xi}_2^{N-1} \qquad.
\end{equation}
In the mildly non-linear regime the growth of gaussian initial conditions
by gravitational instability predicts that the $S_N$ are independent of
scale \cite{Fry84,Ber92} for a scale-free spectrum.
For a top-hat filter of the density field and a pure power-law spectrum,
$P(k)\propto k^n$, it can be shown \cite{P80,Gor,JusBouCol,Ber94} that
to lowest non-trivial order in perturbation theory ($\bar{\xi}_2\ll 1$),
\begin{eqnarray}
  S_3 &=& {34\over 7} - (n+3) \label{eqn:s3n} \\
  S_4 &=& {60712\over 1323} - {62\over 3}(n+3) + {7\over 3}(n+3)^2
  \label{eqn:s4n}
\end{eqnarray}
with a very weak dependence on the matter density, $\Omega_0$ \cite{BCHJ}.
These results have been confirmed by N-body simulations
\cite{ColFre91,BJCP,WeiCol,LIIS,JusBouCol,LMMM,Ber94,Lokas,BGE95,Jus95,ColBouHer}
and are known to be insensitive to redshift space distortions
\cite{LIIS,ColBouHer,Hivetal}
on the mildly
non-linear scales where we will be working (see \S\ref{sec:nbody}).

Measurements of the $S_N$ have been performed on several galaxy catalogues
in two and three dimensions;
see e.g.~Table 1 in Hui \& Gaztanaga~\shortcite{HuiGaz} or
Kim \& Strauss~\shortcite{KimStr}.
We expect that the Anglo-Australian Two Degree Field survey
(2dF\footnote{http://meteor.anu.edu.au/$\sim$colless/2dF})
and the Sloan Digital Sky Survey
(SDSS\footnote{http://www.sdss.org})
will provide excellent catalogues for estimation of the $S_N$ for different
sub-samples to high order.
Assuming growth through gravitational instability from initially small
{\it gaussian\/} fluctuations one can use the additional information
contained in the $S_N$ to relate the properties of the observed galaxy
distribution to the fluctuations in the underlying dark matter.
The observations of CMB anisotropies have enhanced our faith in gravity
as the engine of growth.  The extra assumption required in this step is
that the initial conditions were gaussian.

As with other assumptions in the standard paradigm, our assumption of
gaussianity for the initial fluctuations should be tested against observations.
One difficulty with testing for non-gaussianity has been the lack of a
simple, predictive theory which describes what form, out of the infinite
possibilities, the non-gaussianity should take.
While inflationary models generically predict gaussian fluctuations, not all
inflationary models are ``generic'' in this sense.
And of course there are other, non-inflationary, models of structure formation
which predict non-gaussian perturbations.
The best known of these non-gaussian theories are models based on topological
defects.  However they have problems fitting the observational data and are
difficult to model properly requiring expensive simulations even for the
initial conditions.

Moments have been predicted for a variety of non-gaussian models under
various approximations.
Results for the unsmoothed moments $S_3$ and $S_4$ for arbitrary initial
conditions in perturbation theory have been presented by \cite{FrySch,ChoBou}.
Approximations to defect models have been studied by
\cite{Jaffe,GazMah,GazFos}.  The latter authors also discussed an
approximate calculation of the low-order moments for the model of
Peebles~\shortcite{PeeA,PeeB}, which shall discuss further below.
Weinberg \& Cole~\shortcite{WeiCol} studied models obtained from non-linear
mappings of initially Gaussian fields.
Perhaps the closest ancestor to this work, however, is that of \cite{Coletal},
who used N-body simulations to calculate moments for a variety of non-gaussian
models, including a $\chi_1^2$ model similar to that developed below.

In this paper we investigate the predictions for the lowest moments,
$S_3$ and $S_4$, in a simple, parameterized, non-gaussian ``model''.
The model has useful interpolating properties.  It has one parameter, $\nu$.
The limit $\nu\to\infty$ recovers the gaussian result, with lower values of
$\nu$ being progressively more non-gaussian.  Low values of $\nu$ mimic
cosmological models with rare, high-amplitude peaks.

The outline of the paper is as follows: in the next section we discuss the
non-gaussian model in more detail, in \S\ref{sec:nbody} we discuss
our technique for calculating the moments $S_3$ and $S_4$ using an
N-body code, plus the tests we have run.
In \S\ref{sec:results} we summarize the results,
and in \S\ref{sec:conclusions} we discuss the implications.

\section{The Non-Gaussian Model} \label{sec:model}

The {\sl COBE\/} team \cite{KBBGHSW} first introduced a non-gaussian model
where the real and imaginary parts of the Fourier- (or in their case $\ell$-)
space perturbations, are drawn from independent $\chi^2_\nu$ distributions
with $\nu$ degrees of freedom, adjusted to have zero mean and scaled 
to give the right power spectrum.
While the precise physical significance of this model is unclear, it serves
as a well defined reference with useful interpolating properties, as the
$\chi_\nu^2$ distribution becomes more gaussian as $\nu\to\infty$.
It was shown in \cite{KBBGHSW} that the {\sl COBE\/} data prefer
$\nu\to\infty$, with the gaussian model being 5 times more likely than any
other model tested.

We have modified this model to specify $\chi^2_\nu$ distributions in real
rather than $k$-space.  We proceeded as follows:
Let $\phi_i(x)$ be independent, gaussian random fields of zero mean.
Define, for integer $\nu$,
\begin{equation}
  \psi \equiv {1\over\nu} \sum_{i=1}^{\nu} 
  \left(\phi_i^2 - \left\langle \phi_i^2 \right\rangle \right)
  \qquad .
\end{equation}
Then $\psi$ will be $\chi^2_\nu$ distributed, in real-space, with zero mean.
For $\nu\to\infty$ the fluctuations will be gaussian.  As $\nu$ is lowered
the model becomes progressively more non-gaussian, with more ``rare'' peaks.
The limiting case $\nu=1$ has the perturbations quadratic in a gaussian field,
much as in recently proposed non-gaussian inflation models
\cite{LinMuk,AntMazMot,PeeA,PeeB}.
The reduced moments of the initial conditions are easily calculated
for a $\chi^2_\nu$ distribution.  The mean is $\nu$, and
\begin{eqnarray}
\left\langle (\chi^2-\nu)^2\right\rangle &=& 2\nu, \\
\left\langle (\chi^2-\nu)^3\right\rangle &=& 8\nu, \\
\left\langle (\chi^2-\nu)^4\right\rangle &=& 48\nu+3(2\nu)^2 \qquad .
\end{eqnarray}

\begin{figure}
\begin{center}
\leavevmode
\epsfxsize=8cm \epsfbox{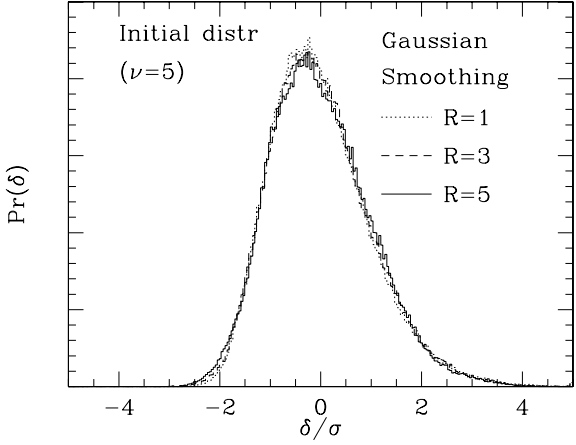}
\end{center}
\caption{The distribution of the initial density fluctuations for $\nu=5$,
smoothed with a gaussian filter of $R=1$, 3 and 5 grid cells in a $64^3$
simulation.  Note that the distribution is skew on all scales.}
\label{fig:icdist}
\end{figure}

A final feature of this model, not shared by other proposals in the literature
which involve non-linear mappings of Gaussian density fields, is that the
distribution of fluctuations will remain approximately the same when smoothed
on a variety of length scales.
This is related to the scaling of this model:
$\left\langle \psi^n\right\rangle$ scales as
$\left\langle\psi^2\right\rangle^{n/2}$.
We show this feature explicitly for the case $\nu=5$ in Fig.~\ref{fig:icdist},
where the distribution of initial fluctuations from one of our simulations
is given for different smoothing scales, spanning a factor of 5 in scale.

It is straightforward to generate initial conditions for this model with
a power-law spectrum.
For $i=1,\cdots,\nu$ we generate independent realizations of a gaussian field
$\phi_i(x)$, with correlation function $\xi_i(r)$, on a grid in the usual way.
We then square $\phi_i$ and sum over $i$ to obtain $\psi$, which we use as an
initial field to generate displacements using the Zel'dovich approximation as
described below.

The only remaining step is to choose the correlation functions $\xi_i(r)$
so that the final correlation function $\xi_\psi(r)$ has a given form.
Writing $\delta\phi_i^2\equiv \phi_i^2-\left\langle \phi_i^2 \right\rangle$
and using the fact that $\phi_i$ is a gaussian one has
\begin{equation}
 \left\langle\delta\phi_i^2({\bf x}_a)\delta\phi_i^2({\bf x}_b)\right\rangle
  = 2 \xi_\phi^2({\bf x}_a-{\bf x}_b) \qquad .
\end{equation}
Note that since the $\phi_i$ are independent, the correlation functions
for the components of $\psi$ simply add.  For simplicity we choose them
to be equal.

\begin{figure}
\begin{center}
\leavevmode
\epsfxsize=8cm \epsfbox{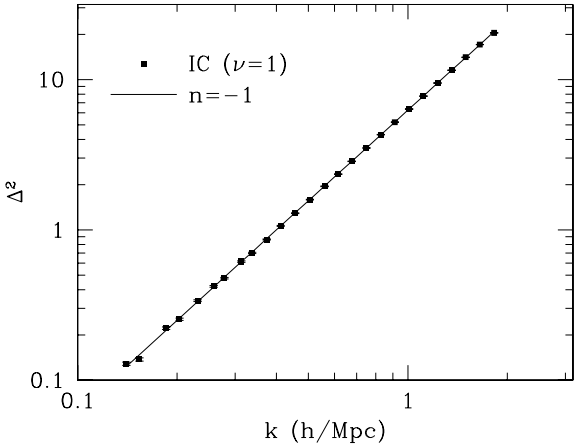}
\end{center}
\caption{The power spectrum of the initial conditions for $\nu=1$ generated
by the method described in the text.  The solid line is the $n=-1$ target
spectrum and the squares indicate the average power spectrum from 50
realizations of the ICs.  The deviation is less than 1\% over more than an
order of magnitude in scale.}
\label{fig:icpk}
\end{figure}

It is important to notice that $\xi_\psi\ge 0$ for all separations, so the
``integral constraint'' (i.e.~the integral of $\xi(r)$ over all space must
vanish) cannot hold for this spectrum.
However we shall be interested in power-law spectra for which $\xi\ge0$
anyway, so this technical point will not affect us.
We shall choose the power law index of the $\phi$ field $n_\phi=-2$ so that
$n_\psi=2n_\phi+3=-1$, close to the slope observed for CDM models on the
scales of interest.
In the absence of finite box size corrections, we get $n_\psi=-1$ by choosing
\begin{equation}
  \Delta_\phi^2(k) = \Delta_{*}^2 \left( {k\over k_*} \right)
  \qquad ,
\end{equation}
where $\Delta_\phi^2\equiv k^3 P_\phi(k)/(2\pi^2)$ is the power spectrum of
the fields $\phi_i$.
This gives $\xi_\phi(r)=(\pi/2)\Delta_{*}^2 (k_*r)^{-1}$ or
\begin{equation}
  \Delta_\psi^2(k) = {2\over\nu} \left({\pi\over 2}\right)^2
  \Delta_{*}^4 \left( {k\over k_*} \right)^2
  \qquad .
\end{equation}
We set $k_*=0.2\,h\,{\rm Mpc}^{-1}$ and adjust $\Delta_*$ so that
$\Delta_\psi^2(k_*)=0.25$, or $\sigma_8\simeq{1\over 2}$.
The same normalization is used in the test simulations described below.

We have checked whether the resulting power spectra for our non-gaussian
models match the $n_\psi=-1$ form we expect.  Since the evolution is a
non-linear function of the initial conditions, it is important to do this
comparison on the initial conditions rather than the evolved spectrum.
We find that there is a departure from a pure power-law at both high-$k$
and low-$k$.
This is to be expected: $\Delta_\psi^2$ is given by an auto-correlation of
$\Delta_\phi^2$, and in the simulations $\Delta_\phi^2$ is missing modes with
wavelengths longer than the box or smaller than the mesh scale.
We have corrected for this effect by modifying the input spectrum of the
$\phi$ field to ensure an initial power-law spectrum for $\Delta^2_\psi$.
While $\Delta^2_\psi$ changes by more than 3 orders of magnitude over the
range of scales we simulate, the correction factor we apply always lies
between 1 and 3 (in power), with the largest correction at low-$k$.
Note that we are not directly modifying the initial power spectrum of the
$\psi$ field here -- it is the convolution integral in the $\phi$ field that
requires this numerical correction.

To sum up:  The procedure outlined above provides an algorithm for generating
realizations of a random field whose 1-point function is a $\chi^2_\nu$
(as has been explicitly checked in the simulations, see \S\ref{sec:results})
and whose 2-point function has a power-law spectrum to within 1\% over more
than an order of magnitude in scale.
We show this explicitly for $\nu=1$ in Fig.~\ref{fig:icpk}.
This initial realization of the $\psi$ field can then be evolved using N-body
simulations to study the effect of gravitational growth on these initial
conditions, as described in the next section.

\section{Calculating the moments} \label{sec:nbody}

We use $N$-body simulations to calculate the moments $S_3$ and $S_4$ in
the quasi-linear regime.
A discussion of several of the relevant numerical issues can be found in
\cite{Jus95,ColBouHer,BGE95,SzaCol,KimStr,JaiBer}.
We use a particle-mesh (PM) code, described in detail in
Meiksin, White \& Peacock~\shortcite{MWP98}.
We have simulated critical density universes ($\Omega_0=1$) with power-law
spectra in boxes of size $150h^{-1}{\rm Mpc}\le L_{\rm box}\le 250h^{-1}$Mpc
on a side, so that the fundamental mode was always well in the linear regime
and the box represented a ``fair'' sample of the universe.
To minimize finite volume effects, we will work on scales less than
about $0.15 R_{\rm box}$, where $L_{\rm box}^3=(4\pi/3)R_{\rm box}^3$
(see below).

We have run simulations with either $64^3$ or $128^3$ particles and a $64^3$
or $128^3$ force ``mesh''.
All the simulations were started at $1+z=20$ to 30 and run to the present
($z=0$).  The evolution was done in log of the scale factor $a=(1+z)^{-1}$.
The time step was dynamically chosen as a small fraction of the inverse square
root of the maximum acceleration, with an upper limit of $\Delta a/a=4$ per
cent per step.
This resulted in a final particle position error of less than 0.1 per cent of
the box size.
As described in Meiksin et al.~\shortcite{MWP98}, the code reproduces the
non-linear power spectrum excellently as compared with semi-analytic fitting
functions (e.g.~Peacock \& Dodds~\shortcite{PD96}) or other N-body codes.
The initial conditions were generated from an initial density field, with
either an $N_{\rm part}$ or $N_{\rm mesh}$ FFT, using the Zel'dovich
approximation.
The particles were initially placed either on a uniform grid, or at random
within cells in a structure with $N_{\rm part}^{1/3}$ cells to a side,
as described in Peacock \& Dodds~\shortcite{PD96}.
Tests indicate the most reliable conditions are when the resolution imposed
by finite particle number matches the force resolution from the mesh and when
the particles were displaced from random positions within cells,
i.e.~the conditions used in the simulations described in
Meiksin et al.~\shortcite{MWP98}.

At the end of each simulation $8,192$ sphere centers were thrown down at
random within the volume and the number of particles in each of multiple
concentric top-hat spheres computed.
The radii of the cells was constrained to be larger than $\simeq 2.5$ (mesh)
cells and smaller than $0.2 R_{\rm box}$.
Since the {\it smallest\/} sphere is more than $5$ grid cells across, the
effects due to the finite resolution of the PM code are minimal.
An explicit check with the higher resolution simulations verified this
expectation, though to achieve per cent level accuracy we need to work
on scales larger than 4 mesh cells (8 in diameter) for the larger boxes
where the grid scale is less non-linear.  This is also to be expected:
in the non-linear regime the structure on the grid scale is determined mainly
by the collapse of larger wavelength modes, which should be well evolved by
the PM code at all times \cite{BouAdaPel}.  Thus as the grid scale becomes
more non-linear, the effects of the finite force grid on perturbations a
few grid cells across are lessened.
We have been careful to ensure that the non-linear scale at $z=0$ has a
wavelength of several mesh zones.
Comparison with different box sizes and resolutions suggested the larger
radius cells, with radii approaching $0.2 R_{\rm box}$, were affected by
the finite box size, with $S_4$ being more affected than $S_3$.
For this reason we restricted the largest cell in the analysis to be less
than $0.15 R_{\rm box}$.
Analytic arguments \cite{HuiGaz} suggest that our boxes are large enough
to avoid finite volume biases at the few per cent level on these scales.

\begin{figure}
\begin{center}
\leavevmode
\epsfxsize=8cm \epsfbox{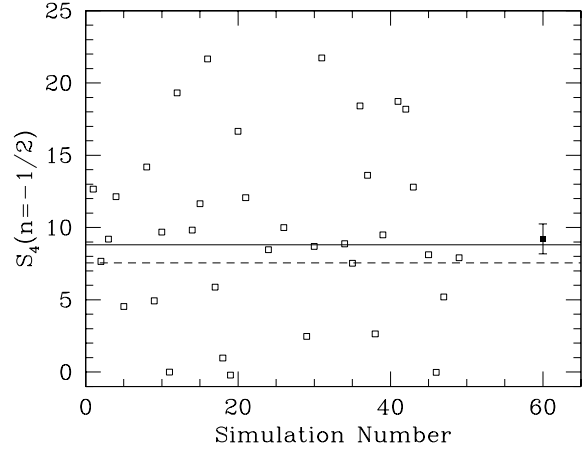}
\end{center}
\caption{Estimates of $S_4(R=10h^{-1}\,{\rm Mpc})$ from the cells thrown for
each of a set of $n=-1/2$ simulations as described in the text.
The average of these values is the dashed line, and the $S_4$ computed from
the entire suite of simulations before averaging is the rightmost (solid)
point with error bar.
This value agrees well with the perturbation theory expectation, shown as the
solid line.}
\label{fig:ncell}
\end{figure}

The moments of the counts-in-cells distribution were calculated and averaged
over many ($>100$) simulations with different realizations of the same initial
spectrum and statistics.  Writing
\begin{equation}
  \mu_M = \left\langle \left( {N-\bar{N}\over\bar{N}} \right)^M \right\rangle
\end{equation}
where $\bar{N}=\langle N\rangle$ and $\langle\cdots\rangle$ represents an
ensemble average (which we approximate by an average over the simulations)
we have \cite{P80}
\begin{eqnarray}
\bar{\xi}_2 &=& \mu_2 - \bar{N}^{-1} \label{eqn:chi22} \\
\bar{\xi}_3 &=& \mu_3 - 3\mu_2\bar{N}^{-1} + 2\bar{N}^{-2} \label{eqn:chi23} \\
\bar{\xi}_4 &=& \mu_4 - 6\mu_3\bar{N}^{-1} - 3\mu_2^2
               +11\mu_2\bar{N}^{-2} - 6\bar{N}^{-3} \label{eqn:chi24}
\qquad .
\end{eqnarray}
Our estimates of $\bar{\xi}_N$ are used to calculate $S_3$ and $S_4$, with
the errors (including correlations) estimated from the run-to-run scatter
and propagated in the usual manner.
Note we average the $\mu_M$ over the simulations and then compute the
$S_N$ rather than averaging the $S_N$ computed from each simulation.
This avoids biases introduced because
$\langle x/y\rangle\ne\langle x\rangle/\langle y\rangle$ \cite{HuiGaz}.
{}From the work of Hui \& Gaztanaga~\shortcite{HuiGaz} we estimate that their
so called ``estimation-biases'' are ${\cal O}(1\%)$ for the cases of interest
here.
We show in Fig.~\ref{fig:ncell} the estimates of $S_4$ from the cells thrown
for a set of simulations with gaussian initial conditions and power-law
index $n=-1/2$.
The average of these values is shown as the dashed line and the $S_4$ computed
{}from the entire suite of simulations as described above is the final (solid)
point with error bar.  This value agrees well with the perturbation theory
expectation -- the solid line -- whereas the naive average does not.
In this Figure the error bars are overestimated for each realization by the
small number of cells thrown, but this error is reduced by the large number
of realization used.

\begin{figure}
\begin{center}
\leavevmode
\epsfxsize=8cm \epsfbox{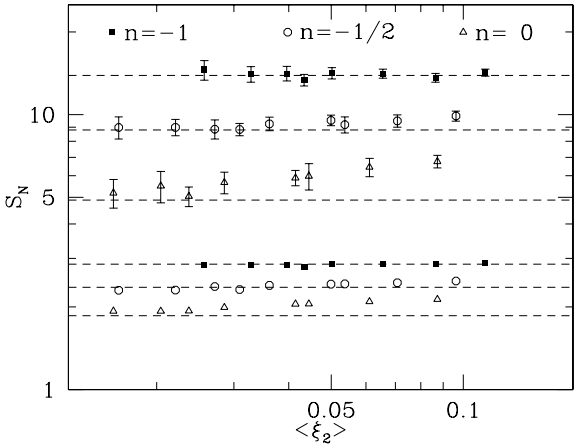}
\end{center}
\caption{The simulation results for $S_3$ and $S_4$ for 3 power-law spectra
with $n=-1$ (solid squares), $n=-1/2$ (open circles) and $n=0$ (open triangles)
as a function of $\bar{\xi}_2$.  The perturbation theory results, valid for
$\bar{\xi}_2\ll 1$, are shown as the horizontal dashed lines.
Error bars (shown for $S_4$) are calculated from the run-to-run scatter in
the simulations.  For $S_3$ the $1\sigma$ error bars are approximately the
same size as the points themselves.
The error bars on the points are correlated.}
\label{fig:test}
\end{figure}

An alternative to the counts-in-cells method described above, is to perform
a maximum likelihood fit to the entire distribution function of the counts
\cite{KimStr}.
This method minimizes errors due to finite volume effects and shot-noise.
However it requires one to know {\it a priori\/} a functional form for the
distribution function.
The Edgeworth expansion, used by Kim \& Strauss~\shortcite{KimStr},
is a valid expansion only near the peak of the distribution.  To remove
unphysical oscillations it must be regularized.
Kim \& Strauss~\shortcite{KimStr} remove the oscillations by convolving the
Edgeworth expansion with a Poisson distribution, to account for the very
sparse sampling they performed.  However once the number of particles in a
cell becomes appreciably greater than unity this convolution no longer
regularizes the distribution and a different expansion is needed.
Typically models for the distribution of counts involve many higher order
moments, beyond $S_4$, making implementation of this method difficult.
For these reasons we have chosen to use the more traditional moments method.
Another alternative to throwing cells \cite{Lokas} is to interpolate the
final density field on a grid, and compute moments of the smoothed field by
summing over the grid points (the smoothing is done using FFT methods).
For our implementation -- using top-hat cells and small grids -- this method,
while significantly faster, was not as accurate.  For fine enough grids and
smooth interpolation (e.g.~CIC or TSC) we would expect to obtain the same
results as the direct counts-in-cells method, but we did not investigate this
in detail.
Finally the method of Szapudi~\shortcite{Istvan} for throwing an effectively
infinite number of cells was too CPU intensive.  For the low order moments we
are considering, the traditional method allows sufficient cells to be thrown
that Szapudi's method is not required.

As a test we first simulated gaussian initial conditions with pure power-law
spectra with $n=-1$, $n=-0.5$ and $n=0$, for which the results are known
analytically (see Eqs.~\ref{eqn:s3n}, \ref{eqn:s4n}) when $\bar{\xi}_2\ll 1$.
The agreement between the numerical and analytic results, shown in
Fig.~\ref{fig:test}, was very good, showing that numerical effects were under
control.
It is interesting to note that for $n\simeq 0$ the perturbation theory results
are obtained only when $\bar{\xi}_2$ is quite small, in contrast to the case
where $n\simeq -1$, for which the perturbation theory results are a good
approximation even for $\bar{\xi}_2\simeq 0.1$.
This agrees with the results of \cite{ColBouHer}, see their Fig.~5.
For $n=-1$ (our fiducial model) the numerical results and analytic predictions
agreed to {\it better\/} than 5 per cent, for both $S_3$ and $S_4$, when
$\bar{\xi}_2\ll 1$.
This was the case for either gaussian initial conditions, or runs with the
non-gaussian model of \S\ref{sec:model} with $\nu\gg 1$.
This is an important result, since it argues that the simulations reported in
\cite{MWP98} can be safely used to estimate correlations between power
spectrum bins in the mildly non-linear regime \cite{MeiWhi}.

\begin{figure}
\begin{center}
\leavevmode
\epsfxsize=8cm \epsfbox{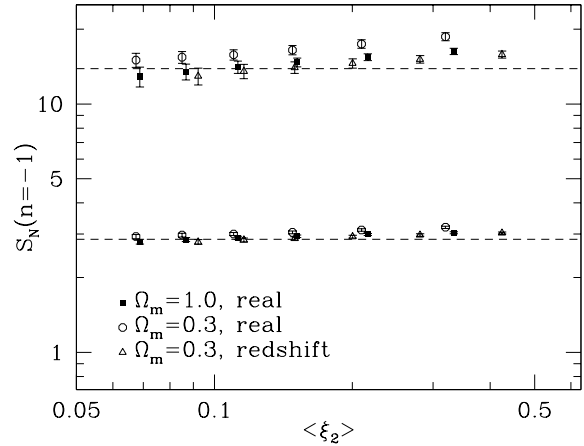}
\end{center}
\caption{The value of $S_3$ and $S_4$ from the $n=-1$ simulations in real
space for a critical density universe (solid squares), an
$\Omega_m=1-\Omega_\Lambda=0.3$ universe (open circles) and in redshift
space in the $\Lambda$ model (open triangles).}
\label{fig:lowomega}
\end{figure}

Having established that the code performs as expected, we simulated models
with $n_\psi=-1$ as described in \S\ref{sec:model} for varying values of
$\nu$.  The results are described in the next section.  Before leaving these
tests however, we show in Fig.~\ref{fig:lowomega} the values of $S_3$ and
$S_4$ for the $n=-1$ spectra with gaussian initial conditions, but varying
$\Omega_m$ and in redshift rather than real space.
Our redshift space results use the distant observer approximation, i.e.~for
each of the simulations we added the $z$-component of the velocity, in units
of the Hubble constant, to the $z$-component of the position before throwing
the cells.
These results confirm that redshift space distortions and variations in
$\Omega_m$ have a negligible impact upon the values of $S_3$ and $S_4$ that
we calculate \cite{LIIS,ColBouHer,Hivetal}.

\section{Results} \label{sec:results}

\begin{figure}
\begin{center}
\leavevmode
\epsfxsize=8cm \epsfbox{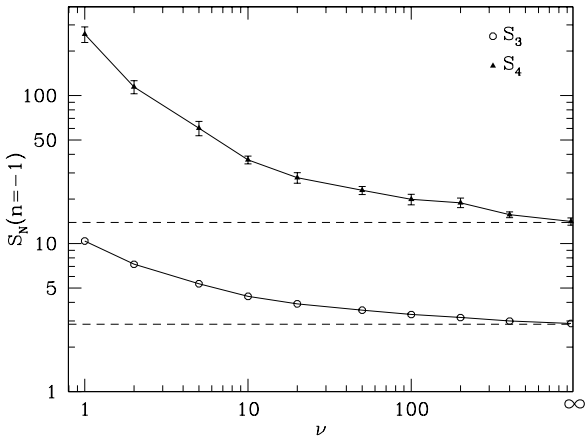}
\end{center}
\caption{The moments, $S_3$ and $S_4$, as a function of $\nu$ for the
non-gaussian model described in \protect\ref{sec:model}.  These values
are evaluated at $10\,h^{-1}{\rm Mpc}$ where $\bar{\xi}_2\simeq 0.1$.
The limit $\nu=1$ has the perturbations quadratic in a gaussian field, as in
recently proposed isocurvature models.  The limit $\nu\to\infty$ (the
rightmost point) recovers the gaussian result.  The predictions of
perturbation theory assuming gaussian initial conditions, $S_3\simeq 2.857$
and $S_4\simeq 13.89$, are shown as the horizontal dashed lines.}
\label{fig:s3s4nu}
\end{figure}

\begin{figure}
\begin{center}
\leavevmode
\epsfxsize=8cm \epsfbox{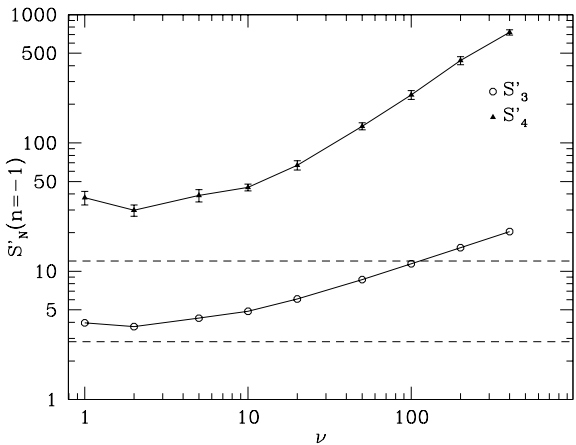}
\end{center}
\caption{The modified moments, $S'_3$ and $S'_4$, defined in the text, as
a function of $\nu$ as in Fig.~\protect{\ref{fig:s3s4nu}}.
The dashed lines show the values $S'_3\simeq 2.82$ and $S'_4=12$ of the
$\chi_\nu^2$ initial conditions.  Note that gravitational instability has
induced a dependence on $\nu$.}
\label{fig:d3d4nu}
\end{figure}

Our main results are shown in Fig.~\ref{fig:s3s4nu}.
Here we plot $S_3(10\,h^{-1}{\rm Mpc})$ and $S_4(10\,h^{-1}{\rm Mpc})$
as a function of $\nu$.  We have chosen this scale because, for our
normalization, $\bar{\xi}_2\simeq 0.1$ at $10\,h^{-1}{\rm Mpc}$.
For $n=-1$ and gaussian initial conditions we would expect $S_3\simeq 2.857$
and $S_4\simeq 13.89$ in the weakly non-linear regime.
These values are shown as the dotted lines in Fig.~\ref{fig:s3s4nu}.
Note that $S_3$ and $S_4$ increase rapidly as the model is made more
non-gaussian.

Once the model is non-gaussian we no longer have reason to expect that
the $S_N$ will be independent of scale.
However non-linear effects process the initial non-gaussianity in a
non-trivial way, as seen in Fig.~\ref{fig:s3s4nu}.
Simple scaling arguments from the initial conditions,
see Eqs.~(\ref{eqn:chi22}-\ref{eqn:chi24}), would suggest that
$\bar{\xi}_3^{\rm init}=\sqrt{8/\nu}\bar{\xi}_2^{3/2}$ and 
$\bar{\xi}_4^{\rm init}=(12/\nu)\bar{\xi}_2^{2}$.
This is to be compared to the gaussian prediction that
$\bar{\xi}_3^{\rm gauss}\propto\bar{\xi}_2^{2}$ and 
$\bar{\xi}_4^{\rm gauss}\propto\bar{\xi}_2^{3}$.
Defining $S'_N$ such that $S'_N$ would be independent of $\nu$ and
$\bar{\xi}_2$ for a $\chi_\nu^2$ field we have
\begin{eqnarray}
S'_3 &\equiv& S_3\ (\bar{\xi}_2\nu)^{1/2} \\
S'_4 &\equiv& S_4\ (\bar{\xi}_2\nu)
\end{eqnarray}
for unsmoothed fields.
For a pure $\chi_\nu^2$ field, $S'_3=\sqrt{8}\simeq 2.82$ and $S'_4=12$.
By calculating the moments of the initial density field from which the
Zel'dovich displacements are generated, we have explicitly checked that
these values are in fact attained in the initial conditions on the grid
(i.e.~without any smoothing).  Any deviation is at the 1 per cent level for
$S'_3$ and $S'_4$ over a wide range of $\nu$.
We show the evolved values of $S'_3$ and $S'_4$ vs $\nu$ in
Fig.~\ref{fig:d3d4nu}, with the gaussian point at $\nu=\infty$ omitted.
Notice that gravity has modified the initial conditions so that the $S'_N$ are
no longer independent of $\nu$.  The curves rise to larger $\nu$, indicating
that the gravitationally induced skewness and kurtosis are a larger fraction
of the initial conditions as the model becomes more gaussian.

In addition, the effects of non-linearity change the scaling with
$\bar{\xi}_2$.  For example, for the $\nu=1$ model the $S_N$ are much less
dependent on $\bar{\xi}_2$ than the $S'_N$, varying by ${\cal O}(10\%)$ for
$0.03\le\bar{\xi}_2\le 0.3$.  This behaviour can be explained as the
gravitational contribution becoming more important than the initial
conditions as clustering evolves, as anticipated by
Fry \& Scherrer~\shortcite{FrySch}.  This result suggests that one must use
care in inferring from the scale-independence of the $S_N$ that the initial
fluctuations are gaussian.  To properly make this inference it may be
necessary to use very large scale or high redshift measurements.

\section{Conclusions} \label{sec:conclusions}

We have used N-body simulations to calculate the low order moments of the
mass density field for a sequence of non-gaussian models with power-law
spectra.
We obtain good agreement with the predictions of perturbation theory for
gaussian initial conditions.
As the real-space perturbations are made progressively more non-gaussian
the moments depart significantly from the values of
Eqs.~(\ref{eqn:s3n},\ref{eqn:s4n}) as shown in Fig.~\ref{fig:s3s4nu}.
Moreover, gravitational evolution modifies the scaling of the $S_N$ with
$\bar{\xi}_2$ and $\nu$.  For example the scale-dependence of the $S_N$
does {\it not\/} follow that of the initial conditions over the range the
simulations probe: $0.03\le\bar{\xi}_2\le0.3$.

\section*{Acknowledgments}

I would like to thank Avery Meiksin and John Peacock for their help in
developing the PM code used in this paper and Lam Hui and Michael Strauss
for useful conversations on measuring $S_3$ and $S_4$ from the simulations.
I am grateful to Pablo Fosalba, Josh Frieman, Bob Scherrer and David Weinberg
for discussions on non-gaussianity, Andrew Liddle for enlightening
conversations on inflationary models and Joanne Cohn for useful comments on
the manuscript.  M.W. is supported by the NSF.


\begin{thebibliography}{99}
\bibitem[\protect\citename{Antoniadis, Mazur \& Mottola }1997]{AntMazMot}
Antoniadis I., Mazur P.O., Mottola E., 1997, Phys. Rev. Lett., 79, 14
\bibitem[\protect\citename{Baugh, Gaztanaga \& Efsatathiou }1995]{BGE95}
Baugh C., Gaztanaga E., Efstathiou G., 1995, MNRAS, 274, 1049
\bibitem[\protect\citename{Bernardeau }1992]{Ber92}
Bernardeau F., 1992, ApJ, 392, 1
\bibitem[\protect\citename{Bernardeau }1994]{Ber94}
Bernardeau F., 1994, Astron. Astrophys., 291, 697
\bibitem[\protect\citename{Bouchet, Adam \& Pellat }1985]{BouAdaPel}
Bouchet F.R., Adam J.C., Pellat R., 1985, Astron. Astrophys., 144, 413
\bibitem[\protect\citename{Bouchet et al. }1992]{BJCP}
Bouchet F.R., Juszkiewicz R., Colombi S., Pellat R., 1992, ApJ, 394, L5
\bibitem[\protect\citename{Bouchet et al. }1993]{BSDFYH}
Bouchet F.R., Strauss M., Davis M., Fisher K., Yahil A., Huchra J., 1993,
  ApJ, 417, 36
\bibitem[\protect\citename{Bouchet et al. }1995]{BCHJ}
Bouchet F.R., Colombi S., Hivon E., Juszkiewicz R., 1995,
  Astron. Astrophys., 296, 575
\bibitem[\protect\citename{Chodorowski \& Bouchet }1996]{ChoBou}
Chodorowski M.J., Bouchet F.R., 1996, MNRAS, 279, 557
\bibitem[\protect\citename{Coles \& Frenk }1991]{ColFre91}
Coles P., Frenk C.P., 1991, MNRAS, 253, 727
\bibitem[\protect\citename{Coles et al. }1993]{Coletal}
Coles P., et al., 1993, MNRAS, 264, 749
\bibitem[\protect\citename{Colley }1997]{Colley}
Colley W.N., 1997, ApJ, 489, 471
\bibitem[\protect\citename{Colombi et al. }1996]{ColBouHer}
Colombi S., Bouchet F.R., Hernquist L., 1996, ApJ, 465, 14
\bibitem[\protect\citename{Feldman, Kaiser \& Peacock }1994]{FKP}
Feldman H., Kaiser N., Peacock J., 1994, ApJ, 426, 23
\bibitem[\protect\citename{Fry }1984]{Fry84}
Fry J., 1984, ApJ, 279, 499
\bibitem[\protect\citename{Fry \& Scherrer }1994]{FrySch}
Fry J., Scherrer R., 1994, 1994, ApJ, 429, 36
\bibitem[\protect\citename{Gaztanaga }1994]{Gaz}
Gaztanaga E., 1994, MNRAS, 268, 913
\bibitem[\protect\citename{Gaztanaga \& Fosalba }1998]{GazFos}
Gaztanaga E., Fosabla P., 1998, preprint [astro-ph/9712263]
\bibitem[\protect\citename{Gaztanaga \& Mahonen }1996]{GazMah}
Gaztanaga E., Mahonen P, 1996, ApJ, 462, L1
\bibitem[\protect\citename{Goroff et al. } 1986]{Gor}
Goroff M.H., Grinstein B., Rey S.-J., Wise M.B., 1986, ApJ, 311, 6
\bibitem[\protect\citename{Heavens }1998]{Heavens}
Heavens A.F., 1998, MNRAS, 299, 805
\bibitem[\protect\citename{Hivon et al. }1995]{Hivetal}
Hivon E., et al., 1995, A\&A, 298, 643
\bibitem[\protect\citename{Hui \& Gaztanaga }1998]{HuiGaz}
Hui L., Gaztanaga E., preprint [astro-ph/9810194]
\bibitem[\protect\citename{Jaffe }1994]{Jaffe}
Jaffe A., 1994, Phys. Rev., D49, 3893
\bibitem[\protect\citename{Jain \& Bertschinger }1998]{JaiBer}
Jain B., Bertschinger E., 1998, ApJ, 509, 517
\bibitem[\protect\citename{Juszkiewicz, Bouchet \& Colombi }1993]{JusBouCol}
Juszkiewicz R., Bouchet F.R., Colombi S., 1993, ApJ, 412, L9
\bibitem[\protect\citename{Juszkiewicz et al. }1995]{Jus95}
Juszkiewicz R., et al., 1995, ApJ, 442, 39
\bibitem[\protect\citename{Kim \& Strauss }1998]{KimStr}
Kim R.S., Strauss M.A., 1998, ApJ, 493, 39
\bibitem[\protect\citename{Kogut et al. }1996]{KBBGHSW}
Kogut A., et al., 1996, ApJ, 464, L29
\bibitem[\protect\citename{Lahav et al. }1993]{LIIS}
Lahav O., Itoh M., Inagaki S., Suto Y., 1993, ApJ, 402, 387
\bibitem[\protect\citename{Linde \& Mukhanov }1997]{LinMuk}
Linde A., Mukhanov V., 1997, Phys. Rev., D56, 535
\bibitem[\protect\citename{Lokas et al. }1995]{Lokas}
Lokas E.L., Juszkiewics R., Weinberg D., Bouchet F.R., 1995, MNRAS, 274, 730
\bibitem[\protect\citename{Lucchin et al. }1994]{LMMM}
Lucchin F., Matarrese S., Melott A.L., Moscardini L., 1994, ApJ, 422, 430
\bibitem[\protect\citename{Meiksin et al. }1999]{MWP98}
Meiksin A., White M., Peacock J., 1999, MNRAS, 304, 851
\bibitem[\protect\citename{Meiksin \& White }1999]{MeiWhi}
Meiksin A., White M., 1999, MNRAS in press [astro-ph/9812129]
\bibitem[\protect\citename{Nusser, Dekel \& Yahil }1994]{NusDekYah}
Nusser A., Dekel A., Yahil A., 1994, ApJ, 449, 439
\bibitem[\protect\citename{Peacock \& Dodds }1996]{PD96}
Peacock J. A., Dodds S. J., 1996, MNRAS, 280, 19
\bibitem[\protect\citename{Peebles }1980]{P80}
Peebles P.~J.~E., 1980, The Large-Scale Structure of the Universe,
Princeton Univ. Press, Princeton
\bibitem[\protect\citename{Peebles }1999a]{PeeA}
Peebles P.J.E., 1999a, ApJ, 510, 523
\bibitem[\protect\citename{Peebles }1999b]{PeeB}
Peebles P.J.E., 1999b, ApJ, 510, 531
\bibitem[\protect\citename{Stirling \& Peacock }1996]{StiPea}
Stirling A.J., Peacock J. A., 1996, MNRAS, 283, L99
\bibitem[\protect\citename{Strauss \& Willick }1995]{StrWil}
Strauss M.A., Willick J.A., 1995, Physics Reports, 261, 271
\bibitem[\protect\citename{Szapudi \& Colombi }1996]{SzaCol}
Szapudi I., Colombi S., 1996, ApJ, 470, 131
\bibitem[\protect\citename{Szapudi }1998]{Istvan}
Szapudi I., 1998, ApJ, 497, 16
\bibitem[\protect\citename{Weinberg \& Cole }1992]{WeiCol}
Weinberg D., Cole S., 1992, MNRAS, 259, 652
\end{thebibliography}
\end{document}